\documentstyle[12pt]{article}
\newcommand{\be}{\begin{equation}}
\newcommand{\ee}{\end{equation}}
\newcommand{\bea}{\begin{eqnarray}}
\newcommand{\eea}{\end{eqnarray}}
\begin{document}
\begin{titlepage}
\begin{flushright}
DTP-MSU-22/97\\
\end{flushright}
\begin{center}
\vspace{10 mm}
{\huge Power--law mass inflation  
in Einstein--Yang--Mills--Higgs black holes}\\
\vspace{8mm}
{\bf D.V. Gal'tsov,}\footnote{Permanent address:  
{\it Department of Theoretical Physics, Physics Faculity, 
Moscow State University, 119899 Moscow, Russia,} email:
galtsov@grg.phys.msu.su}\\
{\it Laboratoire de Gravitation et Cosmologie Relativistes,\\
 Universit\'e Pierre et Marie Curie, 4 place Jussieu \\
 75252 Paris,  France }\\

\vspace{2mm}      
and\\ 

\vspace{2mm}
{\bf E.E. Donets}\\
{\it Laboratory of High Energies, JINR, 141980, Dubna, Russia}
 \end{center}
\vspace{5mm}
\begin{center}
{\large Abstract}
\end{center}
Analytical formulas are presented describing a generic singularity
inside the static spherically symmetric black holes in the 
$SU(2)$ Einstein--Yang--Mills--Higgs theories
with triplet or doublet Higgs field. The singularity 
is spacelike and  exhibits a `power--low mass inflation'.
Alternatively this asymptotic may be interpreted as a pointlike 
singularity with a non--vanishing shear in the Kantowski--Sachs 
anisotropic cosmology.

\vspace{10mm}
PASC numbers: 04.20.Jb, 11.15 Kc, 97.60 Lf.
\end{titlepage}
\newpage

Nature of singularity in `generic' black holes remains one of the unsettled
issues in General Relativity. Exact solutions available demonstrate
possibility of both spacelike (Schwarzschild) and timelike
(Reissner--Nordstr\"om and Kerr) strong singularities, while 
the mass--inflation scenario \cite{minfl} predicts strong spacelike, 
null (weak or strong), or combination of the two. Recently
new insights into the problem were gained in the study of 
non--abelian black holes \cite{dgz,gdz}. Such black holes are known to be 
endowed with external hair violating the naive no--hair conjecture
\cite{vgkb}. Similar 
hair exists {\it inside} the event horizon and may play crucial role
in determining the structure of singularity. In the framework of 
the Einstein--Yang--Mills (EYM) theory a generic singularity inside a
static spherical black hole was shown to be spacelike and of the 
oscillatory nature \cite{dgz}. The singularity is dominated by
both kinetic and `potential' (arising due to dimensional reduction) 
Yang--Mills matter terms. In the case
of the dilatonic version of the same theory (EYMD) a generic black  hole
singularity  was shown to be dominated by the dilaton \cite{dgz}. The
corresponding mass function diverges in the singularity according
to the power--law. Here
we want to show that the same is true for the typical gravity coupled gauge 
theories such as $SU(2)$ Einstein--Yang--Mills--Higgs (EYMH) models with 
triplet and doublet Higgs. The singularity in generic static spherical black 
holes in these theories is strong, spacelike, and dominated by the kinetic
terms of the scalar fields. It can be interpreted as {\it power--law} 
homogeneous mass--inflationary singularity, or, alternatively, as a pointlike 
sigularity of the corresponding Kantowski--Sachs anisotropic cosmological 
model. Our analysis is purely analytical. Recently the EYMH system with 
triplet Higgs was studied numerically \cite{blm}, but the conclusion 
made about an {\it exponential} mass inflation is incorrect.

We consider the $SU(2)$ EYMH theory 
\be
S = \frac{1}{16\pi} \, \int \,  \left(- R - F^2
+ 2|D \Phi|^2 - \frac{\lambda}{2} (|\Phi|^2-\eta^2)^2\right)\sqrt{-g}\,d^4x \, ,
\end{equation} %(1)
\noindent
where $F$ is the YM field strength,  $\Phi$ is a Higgs field in
either vector (real triplet) or fundamental (complex doublet)
representation, $D\Phi$ is the corresponding YM covariant
derivative (in the doublet case $|D \Phi|^2=(D \Phi)^\dagger (D \Phi),\,
|\Phi|^2=\Phi^\dagger \Phi$) and without loss of generality both 
the Planck mass and the gauge coupling constant are set to unity. 
In the flat space--time the triplet version of the theory gives rise 
to regular magnetic monopoles, the doublet version --- to sphalerons.
New physically interesting configurations emerge when gravity is coupled 
in a self--consistent way \cite{gwg}. In particular, static spherical
black holes exist in both cases: monopole  \cite{lnv} and
sphaleron \cite{gmn}. Our aim here is to reveal the {\it generic} 
singularity structure in  static spherically symmetric black hole 
solutions to these theories.

>From rather general considerations it follows that no Cauchy horizons may
form inside generic black holes with non--linear interior hair (although
some non--generic soluions with such horizons still may exists) \cite {dgz}.
Then it is convenient to use the standard curvature coordinates 
\begin{equation}
ds^2=\left(1-\frac{2m(r)}{r}\right)\sigma^2(r) dt^2 - 
\left(1-\frac{2m(r)}{r}\right)^{-1}dr^2
 -r^2 (d\theta^2 + \sin^2 \theta d\varphi^2).
\end{equation} %2
\noindent
taking into account that $2m/r>1$ in the whole interior region. 
 
The purely magnetic Yang--Mills connection relevant to the cases of interest
reads
\begin{equation}
A_{\mu}^a T_a dx^\mu =  (f(r)-1)(L_\varphi d\theta - L_\theta
\sin \theta d\varphi)\, ,
\end{equation} %3
\noindent
where $L_r =T_a n^a$,  $L_\theta=\partial_\theta L_r$,
$L_\varphi= (\sin \theta)^{-1}\partial_\varphi L_r\;,$
$n^a = (\sin \theta \cos \varphi, \sin \theta \sin \varphi, \cos \theta)$, 
$T_a$ are generators of $SU(2)$ in the corresponding representation.
The Higgs field in the triplet case is      
\be %4
\Phi^a T_a =\phi(r) \, L_r,
\ee
while for the doublet  $\Phi =\phi(r) v$, where $v$ is
some (here irrelevant) spinor depending only on angle variables. In both
cases $\phi(r)$ is the only real scalar function of the radial variable.

The system of equations following from the action (1) with this ansatze
may be presented as a set of three decoupled equations for
$f, \phi$ and $\Delta=r^2-2mr$:
\bea  %5,6,7
 \left(\frac{\Delta}{r^2}f'\right)'+\frac{\Delta}{r}f' \phi'^2&=&
\frac{1}{2}\frac{\partial V}{\partial f} -Q \frac{f'}{r}, \\
 \left(\frac{\Delta}{r}\right)'+\Delta \phi'^2&=& 1-2V-Q,   \\
 \left(\Delta\phi'\right)'+\Delta r \phi'^3&=&
\frac{\partial V}{\partial \phi} -Q r\phi',
\eea
where $Q=2\Delta f'^2/r^2$, and
\be %8
V\equiv V(f,\phi,r)=\frac{(f^2-1)^2}{2 r^2}+
\frac{\lambda r^2}{8}(\phi^2-\eta^2)^2 + W^2,
\ee
with $W=f\phi$ in the triplet case and $W=(f+1)\phi$ in the doublet 
one. An equation for  $\sigma$ decouples from the system
\be %9
(\ln\sigma)'=\frac{2}{r}f'^2 +r \phi'^2, 
\ee
and can be easily integrated once $f, \phi$ and $\Delta$  
are found.

The point $r=0$ is a degenerate singular point of the system (5-7),
which is a non--autonomous system of the fifth order. 
It can be shown to give rise to four different solution
branches. Three local branches were presented in \cite{dgz} for the EYM
theory (Higgs field can be added rather trivially \cite{blm}), so
we do not give them explicitly here. Physically, two asymptotics 
correspond to the Schwarzschild and 
Reissner--Nordstr\"om (RN) type singularities, while the third looks
like a RN one with imaginary charge \cite{dgz} which can also
be interpreted as describing a `homogeneous mass inflation' (HMI) 
model \cite{pg,abc}. Genericity of local solutions 
may be explored by counting free parameters; from the above three
only RN has a sufficient number of parameters to be locally generic,
but it fails to be globally generic due to the `second quantization',
as discussed in \cite{dgz}. So we look for another local branch
which should contain five free parameters and satisfy an assumption
$\Delta<0$, necessary condition for the absence of Cauchy horizons. 
(Since the solution we are looking for is supposed to be valid only
in the vicinity of $r=0$, this does not garantee 
in general the absence of Cauchy horizons outside this region,
but numerical integration shows that a continuation 
without Cauchy horizons exists indeed for some values of parameters.) 

This fourth local branch can be found as follows.
Assume the solution to be dominated by the scalar gradient
(kinetic) term, then the truncated system can be integrated analytically
resulting in a five--parameter solution family. Substituting such a solution
into the full system, one finds consistency conditions for 
its validity, it turns out that the convergence radius is non--zero.
The assumption made allows one to neglect the right hand side of
the Eqs.~(5-7) retaining the terms at 
the left hand side. Thus, the truncated system reads
\bea  %   10-12
&& \left(\frac{\Delta}{r^2}f'\right)'+\frac{\Delta}{r}f' \phi'^2= 0,\\
&& \left(\frac{\Delta}{r}\right)'+\Delta \phi'^2= 0,   \\
&& \left(\Delta\phi'\right)'+\Delta r \phi'^3=0.
\eea
It can be easily disentangled leading to the following 
decoupled equations for the YM and scalar fields:
\bea   %13-14
&&f''-\frac{f'}{r}=0,\\
&&\phi''+\frac{\phi'}{r}=0.
\eea
>From here the following first integral is obvious, which may be used
to detect the onset of the asymptotic region numerically:
\be  %15
\frac{d}{dr}\left(f' \phi'\right)=0.
\ee

Once $f$ and $\phi$ are found, the metric function then can be obtained 
>from the simple equation
\be     %16
\frac{\Delta'}{\Delta}=\frac{1}{r} - r \phi'^2.
\ee

General solution to the Eqs. (13-14) contains four constant parameters 
$f_0, \phi_0, b, k$:
\bea       %17-18
&& f=f_0+ b r^2,\\
&& \phi=\phi_0+ k\ln r,
\eea
it shows that the Higgs field logarithmically diverges towards
the singularity, while the Yang--Mills function has a finite limit.
Integrating (16) one finds
\be   %19
\Delta=-2 m_0 r^{(1-k^2)},
\ee
with the fifth (positive) constant $m_0$.
Hence, by counting free parameters, this is a generic
solution with non--positive $\Delta$. Now, to find whether
the truncation (10-12) of the full system (5-7) is consistent, one has to
substitute there the solution (17-19) and to check whether the terms at the 
right hand side of Eqs. (5-7) are small indeed with respect to accounted 
terms. One finds the following condition on the parameter $k$:
\be    %20
k^2>1.
\ee
This means that the metric function $\Delta$ diverges at the singularity.
The corresponding mass--function is also divergent according
to the power--law
\be     %21
m=\frac{m_0}{r^{k^2}}.
\ee
>From the Eq. (9) one can see that the behavior of another metric function
$\sigma$ is dominated by the scalar term, explicitely one obtains
\be %22
\sigma=\sigma_0 r^{k^2},
\ee
with (positive) constant $\sigma_0$.

This local solution  can be interpreted as exhibiting a 
`power--law mass inflation'. 
Usually mass inflation phenomenon is discussed in more general
situation when the stress--energy tensor depends on both
$r$ (time) and $t$ (space)  cordinates in the $T$--region
inside the horizon. A simplified model known as `homogeneous
mass inflation' \cite{pg,abc} deals with space--independent 
matter distributions,
{\it i.e.} depending only on $r$, as in our case. Note that our
definition of the mass function is different from the standard
definition in the mass inflation scenario where the RN--like
contrubution due to the charge is  singled out explicitly.  
In our case there is no Coulomb component of the YM field,
so such contribution is absent. Also, we have assumed
>from the very beginning that no Cauchy horizon form and 
therefore $\Delta(0)\leq 0$ (with possible equality only in 
the singularity). Hence the $1/r$ behavior of the
mass function would mean the RN type asymptotic with imaginary
charge \cite{dgz}. As me already noted, such solution was encountered 
indeed within the context of the mass inflation (HMI solution \cite{abc}). 
However, it was shown that, 
at least in the pure EYM theory \cite{dgz}, HMI interior does not 
correspond to an asymptotically flat exterior spacetime
(no proof has been done so far, however, for the EYMH theory).
In view of (20), the actual divergence of the mass function found here 
is stronger than $1/r$ 
and it is likely that the non--zero treshold $k_0^2$ 
exists below which our generic local interior solution
do not meet asymptotically flat exterior ones. Our analytical solution
seems to agree with numerical results of \cite{blm}, though
we disagree with an interpretation given there as an `exponential inflation
of mass'. Unlike the pure EYM theory, where at some stage of
each oscillation cycle $m(r)$ exhibits a typical for the standard
mass--inflation scenario exponential rise, here one observes
a monotonous power--law behavior.

As it was suggested in \cite{dgz}, another useful interpretation
of the static spherical black hole interiors may be given
in terms of the anisotropic cosmology. Indeed, the metric (2)
in the T--region inside the horizon may be rewritten as a  
Kantowski--Sachs metric
\begin{equation} %23
ds^2=-d\tau^2-X^2(\tau) d\rho^2 -  
 Y^2(\tau)(d\theta^2 + \sin^2 \theta d\varphi^2).
\end{equation} 
The correspondence with our initial  paramatrization (2) is non--pointlike:
\begin{equation}  %24
r=Y,\quad \Delta=-(Y{\dot Y})^2,\quad \sigma=X/{\dot Y},
\end{equation} 
where a dot denotes the derivative with respect to $\tau$
and $t=\rho$.

Field equations for YM and Higgs functions look fairly simple
\bea    %25-26
&& \frac{d}{d\tau}({\dot f}X)=-\frac{X}{2} \frac{\partial V}{\partial f},\\
&& \frac{d}{d\tau}({\dot \phi}XY^2)=-X \frac{\partial V}{\partial \phi},
\eea
where $V$ is given by (8) with $r$ replaced by $Y$.
One of the Einstein equations contains only $Y$ and matter fields: 
\begin{equation} %27
2Y{\ddot Y}+{\dot Y}^2 +1 +2 {\dot f}^2 + {\dot \phi}^2 -2 V=0,
\end{equation} 
while for $X$ the following first order equation holds
\begin{equation} %28
{\dot X}=X{\dot Y}^{-1}\left({\ddot Y} +Y {\dot \phi}^2
+ 2 {\dot f}^2 Y^{-1}\right).
\end{equation} 

The solution above can be obtained neglecting the right sides of
Eqs.~(25,26)  and omitting in the Einstein equations all matter terms except
for those containing ${\dot \phi}$.  
Kantowski--Sachs metric variables in terms of $\tau$ will read 
\begin{equation}  %29
X\sim \tau^{(k^2-1)/(k^2+3)},\quad Y\sim \tau^{2/(k^2+3)}.
\end{equation} 
With account for (20), this corresponds to a pointlike  
anisotropic singularity. It is common to characterize cosmological
expansion using an expansion parameter $\theta$ and a shear $\tilde\sigma$.
Using (29) one obtains
\bea
&&\theta =\frac{{\dot X}}{X}+2 \frac{{\dot Y}}{Y}=\frac{1}{\tau},\\
&&{\tilde\sigma} =\frac{1}{\sqrt{3}}
\left(\frac{{\dot X}}{X} - \frac{{\dot Y}}{Y}\right)=
\frac{k^2-3}{\sqrt{3}(k^2+3)\,\tau}.
\eea

Comparing this power--law mass--inflationary solution with the
corresponding solution to the EYMD system \cite{gdz}, one finds
that they are essentially the same (apart from the different range
of parameters). This is not surprising since in both cases
the main matter contribution to the Einstein equations comes from
the gradient of the scalar field.
However, in the dilaton case $k^2<1$ is allowed, such solutions
were found numerically in \cite{gdz}. This corresponds to
a `cigar' singularity.

Thus, scalar--dominated singularity is rather typical for
non--Abelian gravity coupled theories including scalar fields.
The singularity is spacelike as it is generally beleived. 
Unlike the pure EYM theory it is not oscillating.
Its nature is the most transparent in the Kantowski--Sachs
interpretation where it looks like a point or cigar singularity
usually associated with the `stiff matter' equation of state.
It is easy to check that the scalar kinetic dominance leads indeed
this kind of an effective state equation. Asymptotic similar
to the present one was found by Paul, Datta and Mukherjee
in the simpler case of a single scalar field \cite{pdm}.
Kantowski--Sachs cosmology with dilaton--axion matter 
(without vector fields) as a source was studied recently
by Barrow and Dabrowski \cite{bd}.

Alternative interpretation of the generic EYMH singularity
can be given within the framework of the `mass inflation':
mass function diverges towards the singularity as a negative power
of the (curvature) radial coordinate. 

 D.V.G. thanks the Laboratory of the Gravitation and Cosmology at the
 University P. and M. Curie, Paris, for  hospitality in May
 1997 while this work was in progress.
 Helpful  discussions with J.~D.~Barrow, G.~Cl\'ement, R.~Kerner, and 
 M.~S.~Volkov are gratefully acknowledged. 
 The research was supported in part by the RFBR grants 96--02--18899, 18126.

\end{document}